\documentclass[preprintnumbers,amsmath,amssymb]{revtex4}
\usepackage{graphics}

\begin{document}

\title{Exact Renormalization Group Treatment of the 2D Non Self-Dual Ising Lattices}

\author{Tuncer Kaya}
\email{tkaya@yildiz.edu.tr}
\affiliation{Physics Department, Y{\i}ld{\i}z Technical University, 34220 Davutpa\c sa-Istanbul/Turkey\\
              }


\begin{abstract}
In this work, an exact renormalization group treatment of honeycomb lattice leading to an exact relation between the coupling strengths of the honeycomb and the triangular lattices is presented. Using the honeycomb and the triangular duality relation, the critical coupling values of honeycomb and triangular lattices are calculated exactly by the simultaneous solution of the renormalized relation and the duality relation, without using the so-called star-triangular transformation. Apparently, the obtained coupling relation is unique.  It, not only  takes place the role of the star triangular relation, but also it is the only exact relation obtained from renormalization group theory other than the 1D Ising chain. An exact pair correlation functions expression relating the nearest neighbors and the next nearest neighbor correlation functions are also obtained for the honeycomb lattice. Utilizing this correlation relation, an exact expression of the correlation length of the honeycomb lattice is calculated analytically for the coupling constant values less than the critical value in the realm of the scaling theory. The critical exponents $\nu$ and $\alpha$ are also calculated as $\nu=1$ and $\alpha=0$. 

\end{abstract}
 
\maketitle

\section{\label{sec:level1}INTRODUCTION}
The Ising model was  introduced by Lenz and initially studied by Ising \cite{Ising} in one dimension (1D). Following the proof of the existence of phase transition for this model in 2D by Peierls \cite{Peierls}, it has been subject of extensive activity. The prediction of the critical coupling strength of the 2D square lattice by Kramers and Wannier \cite{Kramer}, by using the duality transformation, is an important step in the investigation of the self-dual square lattice Ising model. Three years later, Onsager \cite{Onsager} and Kaufman \cite{Kaufman,Kaufman1}  computed the free energy of the square lattice, in the absence of external field, by the transfer matrix method. Later, an exact solution in the presence of external field was obtained by Yang \cite{Yang}. The obtained results in their papers are considered as the most important development and progress in the investigation of the Ising model. The investigation of non self-dual lattices such as honeycomb and triangular lattices with transfer matrix method are impossible. On the other hand, the application of the duality transformation to the non self-dual lattices is complicated by the presence of non self-duality. To overcome this complication, Wannier wrote down explicitly the so-called star-triangular transformation \cite{Wannier} to obtain the corresponding exact critical coupling strength values of those lattices. One can find more information in the classical text of McCoy and  Wu \cite{McCoy}.

 As seen from these monumental works, the transfer matrix method and the method of duality transformation play the central role in the investigation of the Ising model. The impact of these monumental works can not be overestimated. Quite apart from explicitly demostrating the existence of phase transitions, they established a benchmark for simulations, motivated the search for exact solution of other models and has served as a testing ground for the efficiency of new theoretical ideas. 

This paper, however, focuses on one of the main stream approaches to the study of Ising model: the real space renormalization group (RSRG) theory with the hope  to obtain some new exact result in this picture.
The RSRG approach to the theory of phase transitions is based on ideas first propounded by Kadanoff \cite{Kadanoff} and subsequently developed by Wilson \cite{Wilson}. The fundamental idea of the RSRG approach to critical phenomena is the calculate the partition function of a system by successively thinning out its degrees of freedom. To this end an iteration procedure is set up where in each step, a certain fraction of the variables of the system is summed up.  The RSRG approach, therefore,  is assumed hypothetically to be an exact theoretical scheme. The application of this seemingly simple method to the Ising system, however, creates some complications in obtaining the renormalized coupling parameters and no further progress can be made without introducing some sort of  approximation or truncation of some terms appearing in the decimated partition function. Since RSRG method is used close to  half a  century  in the treatment of the Ising model, it is difficult to give an overall view of all the different approaches in a reseach paper. But, there are many text books \cite{Nigel,Cardy,Binney} and review papers \cite{Plascak,Efrati} on this subject. Therefore, it is proper to just focus on the results of the applications of this approach to the Ising model in general. 

The application of the RSRG approach to the 1D Ising chain reproduces the previously obtained exact solution since the decimation transformation of the 1D Ising chain leaves the form of Hamiltonian invariant. This means that no new interactions are produced by the decimation process: the nearest-neighbor interactions are retained, but with a renormalized interaction as the sole effect of the decimation. The corresponding calculations for the 2D square lattice \cite{Kadanoff1} and the 3D cubic Lattice prove \cite{Kaya} to be more typical of the RSRG method than the 1D counterpart. In these examples, approximations to the decimated Hamiltonian are necessary to obtain a tractable computational scheme. In other words, the RSRG theory produces approximate results for the Ising model except for the 1D Ising chain. Neverheless, the obtained results go beyond mean field theory \cite{Bragg,Wysin,Kaya1} in several respects and can be systematically improved \cite{Hilhorst}.

At this point, it is important to point out once more that the RSRG theory has an exact solution only for the 1D Ising chain. Therefore, a new exact result obtained from the RSRG approach can be considered valuable and important. In fact, the purpose and motivation of this paper is to present the exact result obtained recently by us for the honeycomb lattice from the RSRG approach in a rigorous manner. Here, {\it{the rigorous manner}} means that we do not produce new concepts or conjectures in the derivation of the exact result. 
Instead we simply use the known RSRG approach. First, the decimation transformation is going to be applied to the original honeycomb lattice. Furthermore, we are going to  try to express the decimated Hamiltonian exactly. It is well-known that the decimation transformation of the honeycomb lattice produces the decimated triangular structure. Writing an Hamiltonian for the decimated lattice properly enables us to obtain an interrelation between the coupling strengths of the original and the decimated lattices. Luckily, as we are going to see, the decimated Hamiltonian is expressed exactly with  only pairwise interaction terms, which are equal to each other for all possible configurations. Of course, the obtained exact relation includes two different coupling strength parameters which makes it impossible to determine the values of the coupling strengths of these lattices. To overcome this difficulty, one needs another relation between the coupling strengths of these lattices. To this end, the exact relation $\tanh\kappa_{t}^{c}=e^{-2\kappa_{h}^{c}}$ obtained from the honeycomb and triangular duality treatment \cite{Kramer,Baxter}, can be used. Here $\kappa_{h}^c$ and $\kappa_{t}^{c}$ denote the critical coupling strengths of honeycomb and triangular lattices respectively. 
The simultaneuos solution of these two equation produces the exact coupling strength values of the honeycomb and triangular lattices. It is important to notice that conventionaly, it is necessary to use  the so called-star-triangular transformation \cite{Wannier,Baxter,pathria} to obtain the critical coupling strength values. 
In other words, the  coupling strength relation obtained from the RSRG approach plays the role of the star-triangular transformation in calculating the exact values of the coupling strength of these lattices. Furthermore, as we are going to see,  one can obtain an exact pairwise correlation function relation between the nearest neighbors and the next nearest neighbors from the exactly obtained decimated partition function. From this correlation function relation, one can calculate  the correlation length of the honeycomb lattice, in the realm of scaling theory, for the values of coupling strength less than the critical value $\kappa_{h}^{c}$ exactly, as well as obtain  exact analytical correlation function expressions for the honeycomb lattice. If the mathematical difficulty of transfer matrix method \cite{McCoy1,Yamada,Wu} in the derivation of certain correlation function expressions of the square lattices is considered, the mathematical simplicity and clarity in the derivation of the correlation function relations for honeycomb lattice can be considered as an  important progress in the treatment of the Ising model in general. In other words, the star-triangular transformation can be only used for the determination of the coupling strengths of honeycomb and triangular lattices, whereas the  exact renormalization relation obtained here enables us to calculate the correlation functions almost straightforwardly in the realm of scaling theory.

\section{\label{sec:level1}THE RSRG APPROACH TO THE HONEYCOMB LATTICE}
We start with the following partition function in the absence of external field of a honeycomb lattice consisting of $N$ spins, namely
\begin{equation}
Q_{N}=\sum_{\{\sigma_{i}\}}\textstyle{\exp}\left[\displaystyle{\sum_{i=1}^{N/2}}\kappa_{h}\sigma_{0,i}(\sigma_{1,i}+\sigma_{2,i}+\sigma_{3,i})\right].
\end{equation}
Here $\sigma_{0,i}$ is the central spin at the $i^{th}$ site while $\sigma_{z,i}$ are the three nearest neighbors around the central spin. The coupling constant can be expressed  in terms of the nearest neighbor interaction constant $J$ as $\kappa_{h}=J/kT$.  $k$ and $T$ denote the Bolzmann's constant and temperature of the system respectively. The set $\{\sigma_{i}\}$ is the union of the set $\{\sigma_{0,i}\}$ and $\{\sigma_{z,i}\}=\{\sigma_{1,i},\sigma_{2,i},\sigma_{3,i}\}$. $N$ is the total number of spins of the system. Writing the  summand in Eq. (1) as
\begin{equation}
\prod_{i=1}^{N/2}\exp[\kappa_{h}\sigma_{0,i}(\sigma_{1,i}+\sigma_{2,i}+\sigma_{3,i})],
\end{equation}
the summations over $\sigma_{0,i}(=\pm1)$ can be carried out straightforwardly, with the result
\begin{equation}
\prod_{i=1}^{N/2}2\cosh\kappa_{h}(\sigma_{1,i}+\sigma_{2,i}+\sigma_{3,i}).
\end{equation}
Thus, the partition function assumes the form
\begin{equation}
Q_{N}=\sum_{\!\{\sigma_{z,i}\!\}}\textstyle{\exp}\left[\displaystyle{\sum_{i=1}^{N/2}}\ln2\cosh\kappa_{h}(\sigma_{1,i}+\sigma_{2,i}+\sigma_{3,i})\right].
\end{equation}
It is important to notice that the partition function presented by Eq. (1) is written for the  honeycomb lattice of $N$ spins while the partition function, expressed in Eq. (4), has triangular lattice structure of $N/2$ spins. The initial partition function written for honeycomb lattice is called as the original lattice, while the final partition function is named as the decimated partition function. 
The simple mathematics used to obtain the decimated partition function is called decimation transformation. Apparently, the term $\ln2\cosh\kappa_{h}(\sigma_{1,i}+\sigma_{2,i}+\sigma_{3,i})$ is equal to $\ln2+\ln\cosh\kappa_{h}(\sigma_{1,i}+\sigma_{2,i}+\sigma_{3,i})$. The crucial step now consists in expressing Eq. (4) in a form similar to Eq. (1). To this end, if the simple trigonometric relation
$\cosh(x+y)=\cosh(x)\cosh(y)(1+\tanh(x)\tanh(y))$ is used, then the term $I=\ln\cosh\kappa(\sigma_{1,i}+\sigma_{2,i}+\sigma_{3,i})$  can be expressed as
\begin{eqnarray}
&&I\!=\!\ln[\cosh\kappa_{h}(\sigma_{1,i}\!+\!\sigma_{2,i})\cosh\kappa_{h}(\sigma_{3,i})
{}\nonumber\\&&(1\!+\!\tanh\kappa_{h}(\sigma_{1,i}+\sigma_{2,i})\tanh\kappa_{h}(\sigma_{3,i}))].
\end{eqnarray}
Using the following relations, 
\begin{eqnarray} 
&&\ln\cosh\kappa_{h}(\sigma_{1,i}+\sigma_{2,i})=\frac{1+\sigma_{1,i}\sigma_{2,i}}{2}\ln\cosh2\kappa_{h},
{}\nonumber\\&&
\tanh\kappa_{h}(\sigma_{1,i}+\sigma_{2,i})=\frac{\sigma_{1,i}+\sigma_{2,i}}{2}\tanh 2\kappa_{h},
{}\nonumber\\&&
\cosh\kappa_{h}\sigma_{3,i}=\cosh\kappa_{h},
{}\nonumber\\&&
\tanh\kappa_{h}\sigma_{3,i}=\sigma_{3,i}\tanh\kappa_{h},
\end{eqnarray}
$I$ can be written more properly as
\begin{eqnarray}
&&I=\ln\cosh\kappa_{h}+\frac{1+\sigma_{1,i}\sigma_{2,i}}{2}\ln\cosh(2\kappa_{h})+\ln[1+\frac{(\sigma_{1,i}+\sigma_{2,i})\sigma_{3,i}}{2}\tanh(2\kappa_{h})\tanh\kappa_{h}].
\end{eqnarray}  

Clearly, we have not been able to establish an exact correspondence between the transformed system and the original one, due to the last term in the Eq.(7). It seems more reasonable now that writing the last term in the form including only the pairwise interaction can be an important step. For this purpose, we are going to consider the series expansion of the function $\ln(1+x)=\sum(-1)^{n+1}\frac{x^{n}}{n}$ for $x<1$. 
Here the sum runs from $1$ to infinity. Apparently, in our case $x$ is equal to $\frac{(\sigma_{1,i}+\sigma_{2,i})\sigma_{3,i}}{2}\tanh(2\kappa_{h})\tanh\kappa_{h}$. The following relations 
\begin{equation}
[\sigma_{3,i}(\sigma_{1,i}+\sigma_{2,i})]^{n}\!= \!\left\{\begin{array}{ll} 2^{n-1}\sigma_{3,i}(\sigma_{1,i}\!+\!\sigma_{2,i}) &\!\!\textrm{for}\:\:\:n \!=\!2k\!+\!1\\
2^{n-1}(1+\sigma_{1,i}\sigma_{2,i})&\textrm{for}\:\:\:n=2k
\end{array} \right.
\end{equation}
can be very useful in the evaluation of the term, where, $k$ runs from zero to infinity. Now, exploiting the relations presented in Eq. (8), this term can be written as,
\begin{eqnarray}
&&\ln[1+\frac{(\sigma_{1,i}+\sigma_{2,i})\sigma_{3,i}}{2}\tanh(2\kappa_{h})\tanh\kappa_{h}]=
\frac{1}{2}\sigma_{3,i}(\sigma_{1,i}+\sigma_{2,i})\sum_{k=0}^{\infty}\frac{(\tanh 2\kappa_{h}\tanh\kappa_{h})^{2k+1}}{2k+1}
{}\nonumber\\&&-\frac{1}{2}(1+\sigma_{1,i}\sigma_{2,i})\sum_{k=1}^{\infty}\frac{(\tanh 2\kappa_{h}\tanh\kappa_{h})^{2k}}{2k}.
\end{eqnarray}
The sums in this relation can be evaluated readily with the following relations,  
\begin{eqnarray}
&&\ln\frac{1+y}{1-y}=2\sum_{k=0}^{\infty}\frac{y^{2k+1}}{2k+1},{}\nonumber\\&&
\ln(1+y)(1-y)=-2\sum_{k=1}^{\infty}\frac{y^{2k}}{2k}.
\end{eqnarray}

Thus, Eq. (9) can be written as
\begin{eqnarray}
&&\ln[1+\frac{(\sigma_{1,i}+\sigma_{2,i})\sigma_{3,i}}{2}\tanh(2\kappa_{h})\tanh\kappa_{h}]=\frac{1}{4}\sigma_{3,i}(\sigma_{1,i}+\sigma_{2,i})\ln\frac{1+\tanh 2\kappa_{h}\tanh\kappa_{h}}{1-\tanh 2\kappa_{h}\kappa_{h}}
{}\nonumber\\&&+\frac{1}{4}(1+\sigma_{1,i}\sigma_{2,i})\ln[1-(\tanh 2\kappa_{h}\tanh\kappa_{h})^2].
\end{eqnarray}
Noticing  the equivalence of the sums $\sum\sigma_{1,i}\sigma_{2,i}=\sum\sigma_{1,i}\sigma_{3,i}=\sum\sigma_{2,i}\sigma_{3,i}$ for all configurations,  Eq. (7) can be written as
\begin{eqnarray}
&&\!\!\!I\!\!=\!\!\ln[\cosh\kappa_{h}(\cosh2\kappa_{h})^{1/2}]\!\!+\!\! \frac{1}{4}\ln[1\!\!-\!\!(\tanh2\kappa_{h}\!\tanh\kappa_{h})^2]+
\frac{1}{2}\sigma_{1,i}\sigma_{2,i}\{\ln[\cosh(2\kappa_{h})\frac{(1\!+\!\tanh(2\kappa_{h}\tanh\kappa_{h}))^{3/2}}{(1\!-\!\tanh(2\kappa_{h}\!\tanh\kappa_{h}))^{1/2}}\!]\}.
\end{eqnarray} 
Thus, Eq. (4) can be rewritten as
\begin{equation}
Q_{N}=\sum_{\!\{\sigma_{z,i}\!\}}\textstyle{\exp}\left[\displaystyle{\sum_{i=1}^{N/2}}(\ln2+I))\right].
\end{equation}
Expressing $\ln2+I=I_{r}+I_{s}$, here $I_{r}$ and $I_{s}$ can be defined as
\begin{eqnarray}
&&\!I_{r}\!\!=\!\!\ln[2\cosh\kappa_{h}(\cosh2\kappa_{h})^{1/2}]\!\!+\!\! \frac{1}{4}\ln[1\!-\!(\tanh2\kappa_{h}\tanh\!\kappa_{h})^2]{}\nonumber
\\&&\!I_{s}\!\!=\!\!
\frac{1}{2}\sigma_{1,i}\sigma_{2,i}\{\!\ln[\cosh(2\kappa_{h})\frac{(1+\tanh(2\kappa_{h}\!\tanh\kappa_{h}))^{3/2}}{(1-\tanh(2\kappa_{h}\!\tanh\kappa_{h}))^{1/2}}\!]\}
\end{eqnarray} 
where, the term $\sigma_{1,i}\sigma_{2,i}$ is the next nearest neighbor interaction in the original honeycomb lattice or the nearest neighbor interaction in the decimated triangular lattice. It is important to notice that the function $I_r$ is an analytical function of $\kappa$ and hence play no direct role in determining the critical behavior of the system. On the other hand, the function $I_s$ is the relation that determines the singular part of the system and hence the critical behavior. Thus, it is relevant to define the following partition function,
\begin{equation}
Q_{N/2}=\sum_{\!\{\sigma_{z,i}\!\}}\textstyle{\exp}\left[\displaystyle{\sum_{i=1}^{N/2}} I_{s}\right],
\end{equation}  
to continue the renormalization procedure. Apparently, $Q_N$ can now be expressed as $Q_{N}=\exp(\frac{I_{r}N}{2})Q_{N/2}$. The crucial step turns out to be now to find a partition function which is equivalent to the $Q_{N/2}$. To this end, if the triangular lattice structure of decimated lattice is recalled, an equivalent partition function can be proposed in the following form,
\begin{equation}
Q_{N/2}=\sum_{\{\sigma_{j}\}}\left[\exp\displaystyle\sum_{j=1}^{N/4}\kappa_{t}\sigma_{0,j}(\sigma_{1,j}+\sigma_{2,j}+\dots+\sigma_{6,j})\right],
\end{equation}
here $j$ runs fron $1$ to $N/4$. It is important to notice that $\sigma_{0,j}$ is the central spin at the $j^{th}$ site while the spins $\sigma_{z,j}$,$ z=1, 2,\dots, 6$  are the nearest neighbors of the central spin in the decimated lattice. Apparently, $\kappa_{t}$ denotes the coupling constant of the triangular lattice. Since the partition functions given by Eq. (15) and Eq. (16) are equivalent, one can assert easily that $\kappa_{t}$ must satify the following relation,
\begin{equation}
\frac{1}{2}\ln[\cosh(2\kappa_{h})\frac{(1+\tanh(2\kappa_{h}\tanh\kappa_{h}))^{3/2}}{(1-\tanh(2\kappa_{h}\tanh\kappa_{h}))^{1/2}}]=3\kappa_{t}.
\end{equation}
To obtain the values of the critical coupling strengths $\kappa_{h}$ and $\kappa_{t}$, we need to have another independent relation between the coupling strengths. For this purpose, it is relevant to use the previously obtained honeycomb and triangular duality relation, namely $\tanh\kappa_{t}^{c}=e^{-2\kappa_{h}^{c}}$ \cite{Baxter}. It is important to notice that to obtain the critical values from the duality relation, one needs to use the so-called star-triangular relation \cite{pathria}. Now, considering this duality relation and Eq. (17) simultaneously leads to the following equation,
 \begin{eqnarray}
&&\frac{1}{2}\ln[\cosh(2\kappa_{h}^{c})\frac{(1+\tanh(2\kappa_{h}^{c}\tanh\kappa_{h}^{c}))^{3/2}}{(1-\tanh(2\kappa_{h}^{c}\tanh\kappa_{h}^{c}))^{1/2}}]=3\tanh^{-1}(e^{-2\kappa_{h}^{c}}).
\end{eqnarray}
From this equation the value of the critical coupling strength of honeycomb lattice is calculated as $\kappa_{h}^{c}=0.658479$ which is equal to the exact result $\kappa_{h}^{c}=\frac{1}{2}\sinh^{-1 }\sqrt{3}$. The value of $\kappa_{t}^{c}$ can be calculated straightforwardly as $\kappa_{t}^{c}=0.274653$, which is equal to the exact value $\kappa_{t}^{c}=\frac{1}{2}\sinh^{-1}(\frac{1}{\sqrt{3}})$. Of course, obtaining the exact results from the renomalization approach is significant and important since the exact relation expressed with Eq. (18) is the only exact relation obtained from the renormalization theory except for the simple 1D Ising chain. This means that we have achieved to obtain an exact analytic relation from the RSRG theory after almost a half century. Furthermore, besides this very important result,  we are going to see that the correlation functions of honeycomb lattice are going to be obtained in the realm of the scaling theory in the following section.
\section{\label{sec:level1}THE CORRELATION FUNCTIONS OF THE HONEYCOMB ISING LATTICE}
The exact determination of correlation functions of non self-dual Ising lattices are unsolved problems of statistical physics. Indeed, the exact determination of the correlation functions of the 2D square lattice are possible only between some particular lattice points due to the mathematically cumbersome transfer matrix method. As pointed out earlier, the applications of the transfer matrix method to the non self-dual lattices have not yet been developed . It is, therefore, necessary and also relevant to consider the correlations functions in their scaling form as follows,
\begin{equation}
<\sigma(\vec{r}_i)\sigma(\vec{r}_j)>=\frac{A}{|\vec{r}_{j}-\vec{r}_{i}|^{(d-2+\eta)}}e^{-\frac{|\vec{r}_{j}-\vec{r}_{i}|}{\xi(\kappa)}}.
\end{equation}
where, the critical exponent $\eta=\frac{1}{4}$ is  for the 2D Ising lattices and $r=|\vec{r}_{j}-\vec{r}_{i}|$ is the distance between the $j^{th}$ and $i^{th}$ sites. 
$\xi (\kappa)$ is the correlation length which diverges (or goes to infinity) at the critical point $\kappa_{c}$.  Now, taking the logarithm of both sides of the Eq. (13), and then calculating the derivatives with respect to $\kappa_{h}$ leads to,
\begin{eqnarray}
\!\!\!\!\!\!\!\!\!\!\!\!\!\!\!\!\!\!\!\!\!\!&&<\!\!\sigma_{0,i}\sigma_{1,i}\!\!>=\frac{\cosh2\kappa\tanh\kappa}{-2+4\cosh2\kappa}+\frac{1}{6}[1+3\tanh2\kappa+\frac{\sinh2\kappa}{-1+2\cosh2\kappa}]<\!\!\sigma_{1,i}\sigma_{2,i}\!\!>.
\end{eqnarray}
Adapting the notations of the Eq. (19) to the notation of Eq. (20), it can be written at the critical point, $\kappa_{h}^{c}=0.6585$ of honeycomb lattice as
\begin{equation}
A=0.1924+0.6959\frac{A}{(3)^{1/8}}.
\end{equation}
Here we used the nearest neighbor distance $a=1$, and next nearest neighbor distance $\sqrt{3}$. Solution of this equation produces $A=0.4892$. It is important to notice that we have just used exact calculation to obtain the value of $A$  for the honeycomb lattice. We think it is important to mention that the value of $A$ for square lattice was obtained approximately as $A\simeq0.645$  by McCoy. 

Now, it is proper to calculate the corresponding critical exponents $\nu$ and $\alpha$. To this end, it is possible to obtain the values of these critical exponents from Eq. (20) numerically using the scaling form, $\xi\sim |\kappa-\kappa_{c}^{-\nu}|$ as $\kappa$ is very close to $\kappa_{c}$. Indeed, we have tried to obtain $\nu$ in this manner. 
At this point, we have concluded that the values of $\nu$ depend strongly on the chosen interval. To overcome this irrelevancy, it is relevant to find a wayout analytical method for the calculations of these exponents. For this purpose, we conjecture  that  it is proper to assume the following form of the two-site correlation function is valid for $\kappa<\kappa_{c}$i 
\begin{equation}
<\sigma(\vec{r}_i)\sigma(\vec{r}_j)>=[\tanh(\gamma_{r}\kappa)]^{r}.
\end{equation}
Here, $\gamma_{r}$ is a parameter which can be related to the known parameter of the scaling form by the use of Eq. (19). Thus, one can write readily as 
\begin{equation}
[\tanh(\gamma_{r}\kappa_{c})]=\left[\frac{A}{r^{\eta}}\right]^{\frac{1}{r}}.
\end{equation}
Thus, it is easily obtained that $\gamma_{r}=\frac{1}{\kappa_{c}}\text{arctanh}\left[\left(\frac{A}{r^{\eta}}\right)^{\frac{1}{r}}\right]$.  From Eq. (19) and Eq. (22) one easily gets  to  
\begin{equation}
\tanh(\gamma_{r}\kappa)=\left[\frac{A}{r^{\eta}}\right]^{\frac{1}{r}}e^{-\frac{1}{\xi}}.
\end{equation}
Now, considering $t=\frac{(\kappa_{c}-\kappa)}{\kappa_{c}}\ll1$, the function  $\tanh(\gamma_{r}\kappa)$ can be written as, 
\begin{equation}
\tanh(\gamma_{r}\kappa)=\tanh\left[(1-t)\text{arctanh}\left(\frac{A}{r^{\eta}}\right)^{\frac{1}{r}}\right].
\end{equation}
Furthermore, Eq. (24) can be written in this case as
\begin{equation}
\tanh\left[(1-t)\text{arctanh}\left(\frac{A}{r^{\eta}}\right)^{\frac{1}{r}}\right]=\left[\frac{A}{r^{\eta}}\right]^{\frac{1}{r}}e^{-\frac{1}{\xi}}.
\end{equation}
Taking logarithm of both sides of  Eq. (26) and expanding it with respect to $t$ around zero produces the following equation
\begin{equation}
\log(a)-\left[\frac{\text{arctanh(a)}}{a}-a\text{arctanh(a)}\right]t=\log(a)-\frac{1}{\xi},
\end{equation}
where $a=\left[\frac{A}{r^{\eta}}\right]^{\frac{1}{r}}$. Thus, the last equation easily leads to $\frac{-1}{\xi}\sim-t $. Recalling the scaling form of the  correlation length $\xi\sim|t|^{-\nu}$, one can easily obtain the result $\nu=1$.  Using the scaling law between $\nu$ and $\alpha$ which is expressed as $d\nu=2-\alpha$, here $d=2$ for the 2D Ising lattices, one can also readily obtain the result $\alpha=0$. It is important to notice that the obtained values $\nu=1$ and $\alpha=0$ are the same as the exact results of the 2D square Ising lattice. Considering the universality assumption which states that the critical exponents depends only the dimension of the Ising system, the obtained critical exponents in this work are considered as the exact results in the realm of the universality assumption. At this point it is worth to mention that the heuristic conjecture presented by Eq. (22) turns out to be a relevant presentation of the correlation functions of the 2D honeycomb lattice. In other words, in this section, we have obtained not only the exact values of the critical exponent $\nu$ and $\alpha$, but also we have tested the consistency of the assumption of universality.
\section{DISCUSSION AND CONCLUSIONS}
In this paper, the non-self dual 2D lattices, namely honeycomb and triangular structures are investigated analytically from the RSRG perspective. Exact 
relations between the coupling strengths of these lattices are obtained. The simultaneous solution of these relations and the previously obtained duality transformation relation leads to the exact critical coupling values which were calculated using the so-called star-triangular relations. We consider the obtained exact relation from the RSRG method is important not only because it replaces the star-triangular relation to obtain the critical values, but also since it is the only exact relation calculated from the RSRG picture, except for the 1D Ising chain. In addition, one can see the importance of the obtained exact relation if he or she considers that even numerically simulating such systems on computers remains a great challenge. Furthermore, an exact correlation functions relation is obtained between the nearest and the next nearest neighbors of the honeycomb lattice. With the help of this relation and the scaling form of the two-site correlation function, the critical exponents $\nu$ and $\alpha$ are calculated analytically as $\nu=1$ and $\alpha=0$. We consider these critical exponent values are exact because of the universality assumption which states that a unique universality class shares the same critical exponents. 
In writing this paper, we tried  not to introduce new assumptions in the treatment of  the RSRG approach. On the other hand, a heuristic conjecture is made for the form of the two-site correlation functions to calculate the critical exponents analytically. Luckily, the proposed form is turned out to be quite relevant picture to calculate the critical exponents as well as the two-site correlation functions of the honeycomb lattice for the values $\kappa<\kappa_{c}$.

\end{document}